\documentstyle[11pt]{article} 
\input{amssym.def} 
\def\RR{{\Bbb R}} \def\CC{{\Bbb C}} \def\QQ{{\Bbb Q}}
\def\PP{{\Bbb P}} \def\ZZ{{\Bbb Z}}  
\def\deca{{\sc DEC Alpha 3000}}		% machines tabulated
\def\sparc{{\sc Sun Sparc 20}} 
\def\MapleV{{\sc Maple~V}}			% programs
\def\qq{{\dot{q}}}				%% quaternions
\def\qt{{\dot{t}}}
\def\qx{{\dot{x}}}
\def\qy{{\dot{y}}}
\def\qz{{\dot{z}}}

\def\conv{{\rm Conv}}				%% shortcuts
\def\degs{{^{\circ}}}
\def\oK{{\overline K}}
\def\MM{{M \! M}}
\def\MV{{M \! V}}
\def\SO{\mbox{SO}}
\def\supp{{\rm supp}}
\def\vol{{\rm Vol}}

\def\A{{\cal A}}			%% one-letter
\def\B{{\cal B}}
\def\E{{\cal E}}
\def\I{{\cal I}}

\def\hx{{\widehat{x}}}

\newtheorem{theorem}{Theorem}[section]		%% theorems

\newtheorem{corollary}[theorem]{Corollary}
\newtheorem{definition}[theorem]{Definition}
\newtheorem{Example}[theorem]{Example}
\newtheorem{lemma}[theorem]{Lemma}

\newtheorem{algorithm}[theorem]{Algorithm}

	{\begin{trivlist}\item[\hskip\labelsep{\bf #1 }]}%
	{\end{trivlist}}

\newcommand{\proof}{\noindent {\bf Proof} \hspace{.1in}}
\newcommand{\qed}{\hfill $\Box$ \newline}

\def\lid{\mathrel{\mathchoice {\vcenter{\offinterlineskip\halign{\hfil	%% <=
$\displaystyle##$\hfil\cr<\cr\noalign{\vskip1.2pt}=\cr}}}
{\vcenter{\offinterlineskip\halign{\hfil$\textstyle##$\hfil\cr<\cr
\noalign{\vskip1.2pt}=\cr}}}
{\vcenter{\offinterlineskip\halign{\hfil$\scriptstyle##$\hfil\cr<\cr
\noalign{\vskip1pt}=\cr}}}
{\vcenter{\offinterlineskip\halign{\hfil$\scriptscriptstyle##$\hfil\cr
<\cr
\noalign{\vskip0.9pt}=\cr}}}}}
\input{psfig}
 
\begin{document} 
\title{\bf
A General Solver Based on Sparse Resultants
\thanks{
Most of this work was conducted as part of the author's
Ph.D.\ thesis in the Computer Science Division
of U.C.\ Berkeley (completed in 1994).
}}%title
 
\author{Ioannis Z.~Emiris\\
\\ Dept Informatics \& Telecoms, University of Athens, Greece \\
{\tt emiris@di.uoa.gr}}

\maketitle
\thispagestyle{empty}

\begin{abstract}
Sparse elimination exploits the structure of polynomials by
measuring their complexity in terms of Newton polytopes instead
of total degree.
The sparse, or Newton, resultant generalizes
the classical homogeneous resultant and its degree is a function
of the mixed volumes of the Newton polytopes.
We sketch the sparse resultant constructions of Canny and Emiris
and show how they reduce the problem of root-finding to an
eigenproblem.
A novel method for achieving this reduction is presented which
does not increase the dimension of the problem.
Together with an implementation of the sparse resultant construction,
it provides a general solver for polynomial systems.
We discuss the overall implementation and illustrate its use by
applying it to concrete problems from vision, robotics and structural
biology.
The high efficiency and accuracy of the solutions suggest that
sparse elimination may be the method of choice for systems of moderate size.
\end{abstract}

\section{Introduction}

The problem of computing all common zeros of a system of polynomials
is of fundamental importance in a wide variety of scientific
and engineering applications.
This article surveys an efficient method based on the sparse resultant
for computing all {\em isolated}
solutions of an arbitrary system of $n$ polynomials in $n$ unknowns.
In particular, we exploit the algorithms of Canny and
Emiris~\cite{CaEm,EmCa94} for
constructing sparse resultant formulae which yield nontrivial
multiples of the resultant.
We show that the matrices obtained allow the reduction of the
root-finding problem to the eigendecomposition of a square matrix.
The emphasis here is placed on practical issues and the application
of our implementation to concrete problems.

We describe very briefly the main steps in sparse elimination and
the construction of sparse resultant matrices.
Most proofs are omitted but can be found
in~\cite{CaEm,EmCa:issac,EmCa94,Em94,EmiPhd}.
The study of coordinate rings of varieties
in $K^n$, where $K$ is a field,
has been shown to be particularly
useful in studying systems of polynomial equations.
We concentrate on zero-dimensional varieties for which
it is known that the coordinate ring forms a finite-dimensional
vector space and, actually, an algebra over $K$.
An important algorithmic question is the construction of an explicit
monomial $K$-basis for such a space.
Based on monomial bases, we may generate generic endomorphisms or
multiplication maps for any given polynomial,
as outlined in section~\ref{Sspelim}.

Root finding is reduced to an eigenproblem and then existing
techniques are employed from numerical linear algebra.
An important feature of our method is precisely that it reduces to
matrix operations for which relatively powerful and accurate 
implementations already exist.
Section~\ref{Spo_ures} discusses this method in connection to both
resultant algorithms for the case of adding
an extra $u$-polynomial to obtain an overconstrained system.
This is the classical method, used in defining the $u$-resultant;
it possesses the advantage that the matrices have a lot of known structure.

A relatively novel approach that keeps the number of polynomials fixed is
proposed in section~\ref{Spo_hide} where one of the variables is {\em hidden}
in the coefficient field, thus producing an overconstrained system.
We show how the calculation of all isolated roots again reduces to
an eigenproblem.
This technique keeps the number of polynomials fixed, which has been
observed to be important in practice.
On the other hand, it leads to arbitrary matrix polynomials for which
we have to calculate all eigenvalues and eigenvectors.

This approach has been implemented in C by the author and
provides, together with an eigenvalue solver,
a self-contained and fast polynomial solver.
% It is intended to be part of the algebraic toolkit built by the author and
% A.~Rege under the supervision of J.~Canny~\cite{Ca94}.
Section~\ref{Spo_numerical} describes our implementation and
discusses the practical issues that arise thereof.
In particular, we consider issues of numerical stability
and conditioning of the matrices.

Our techniques find their natural application in
problems arising in a variety of fields, including problems expressed
in terms of geometric and kinematic constraints.
As an empirical observation, polynomial systems encountered in
robot and molecular kinematics, motion and aspect ratio calculation in vision
and geometric modeling are characterized by small mixed volume
compared to their Bezout bound.
The complexity of our methods depends directly on this sparse structure,
in contrast to Gr\"obner bases.
Sparse homotopies were proposed in order to exploit the same
structure~\cite{HuSt,VVC94},
yet they still suffer from accuracy problems and the possibility that
some solutions may not be found.
Lastly, resultant-based methods include a large fraction of offline
computation: the first phase, including the construction of the matrix,
has to be executed only once for a given set of supports.
For every specific instance, the coefficients are specialized and
then the online phase has to be executed.

We describe in detail two problems from vision, robot kinematics
and structural biology.
The first problem, analyzed and solved in section~\ref{Sapp_pts2motion},
is a standard problem from photogrammetry.
Given information about a static scene seen by two positions
of a camera, the camera motion must be computed.
When the minimum amount of information is available so that the problem
is solvable, an optimal sparse resultant matrix is constructed and
the numerical answers are computed efficiently and accurately by our
implementation.
Our method exhibits competitive speed, as compared
to previous approaches, and better accuracy.

The second application, in section~\ref{Sapp_mole},
comes from computational biology and reduces to
an inverse kinematics problem.
The symmetry of the molecule at hand explains the high multiplicity of the
common zeros, which leads us to compare the two approaches of defining
an overconstrained system, either by adding a $u$-polynomial or by
hiding one of the input variables.
Both methods are used for different instances in order for all roots to
be calculated accurately.

We conclude with some open questions in section~\ref{concl}.

\section{Sparse Elimination} \label{Sspelim}

Sparse elimination generalizes several results of classical elimination
theory on multivariate polynomial systems of arbitrary degree
by considering the structure
of the given polynomials, namely their Newton polytopes.
This leads to stronger algebraic and combinatorial results in general.
Assume that the number of variables is $n$;
roots in $(\CC^*)^n$ are called {\em toric}.
By concentrating on $\CC^*$ we may, consequently,
extend our scope to {\em Laurent polynomials}.
We use $x^e$ to denote the monomial $x_1^{e_1} \cdots x_n^{e_n}$,
where $e =(e_1,\ldots,e_n)\in \ZZ^n$ is an exponent vector or,
equivalently, an integer lattice point, and $n\in\ZZ_{\geq 1}$.
Let the input Laurent polynomials be
\begin{equation} \label{square_sys}
f_1,\ldots,f_n\in K[x_1,x_1^{-1},\ldots,x_n,x_n^{-1}]=K[x, x^{-1}]
\end{equation}
where $K$ is a field.

Let $\A_i = \supp(f_i) = \{a_{i1},\ldots,a_{i \mu_i}\}\subset \ZZ^n$
denote the set, with cardinality $\mu_i$,
of exponent vectors corresponding to monomials in $f_i$
with nonzero coefficients.
This set is the {\em support} of $f_i$:
$$
f_i = \sum_{a_{ij}\in\A_i} c_{ij} x^{a_{ij}},\qquad c_{ij}\neq 0.
$$

\begin{definition}
The {\em Newton polytope} of $f_i$ is the convex hull of support $\A_i$, denoted
$Q_i=\conv(\A_i) \subset \RR^n$.
\end{definition}

For arbitrary sets in $\RR^n$ there is a natural associative and
commutative addition operation
called Minkowski addition.

\begin{definition}
The {\em Minkowski sum} $A+B$ of sets $A$ and $B$ in
$\RR^n$ is
$$
A+B = \{a+b| \, a\in A, b\in B\}\,\subset\RR^n.
$$
If $A$ and $B$ are convex polytopes then $A+B$ is a convex
polytope.
\end{definition}

Let $\vol(A)$ denote the Lebesgue measure of $A$ in $n$-dimensional euclidean
space, for polytope $A\subset\RR^n$.

\begin{definition} \label{Din_mv1}
Given convex polytopes $A_1,\ldots,A_n\subset\RR^n$, there is a
unique real-valued function
$\MV(A_1,\ldots,A_n)$, called the {\em mixed volume} of $A_1,\ldots,A_n$
which has the following two properties.
First, it is
multilinear with respect to Minkowski addition and scalar multiplication
i.e.\, for $\mu,\rho\in\RR_{\geq 0}$ and convex polytope $A_k'\subset\RR^n$
$$
\MV(A_1,\ldots,\mu A_k+\rho A_k',\ldots,A_n) =
\mu \MV(A_1,\ldots,A_k,\ldots,A_n)+\rho \MV(A_1,\ldots,A_k',\ldots,A_n).
$$
Second,
$$
\MV(A_1,\ldots,A_n)=n! \;\vol(A_1),\qquad\mbox{when }A_1=\cdots=A_n.
$$
\end{definition}

Notationally, we use
$$
\MV(Q_1,\ldots,Q_n)=\MV(\A_1,\ldots,\A_n)=\MV(f_1,\ldots,f_n).
$$
We are now ready to state Bernstein's theorem~\cite{Be}, the
cornerstone of sparse elimination, generalized to arbitrary varieties.

\begin{theorem} \label{Tin_bkk} {\rm \cite[sect.~5.5]{Ful93}}
Given are polynomials $f_1,\ldots,f_n\in K[x, x^{-1}]$ with Newton polytopes
$Q_1,\ldots,Q_n$.
For any isolated common zero $\alpha\in(\CC^*)^n$,
let $i(\alpha)$ denote the intersection multiplicity at this point.
Then $\sum_\alpha i(\alpha) \leq \MV(Q_1,\ldots,Q_n)$, where the sum ranges over all
isolated roots.
Equality holds when all coefficients are generic.
\end{theorem}

Canny and Rojas have substantially weakened the requirements for
equality~\cite{CaRo}.
A recent result extends the bound on non-toric roots.

\begin{theorem} {\rm \cite{LiWa}} \label{Tallroots}
For polynomials $f_1,\ldots,f_n\in\CC[x,x^{-1}]$ with supports
$\A_1,\ldots,\A_n$
the number of common isolated zeros in $\CC^n$, counting multiplicities,
is upwards bounded by $\MV(\A_1 \cup \{0\},\ldots,\A_n\cup\{0\})$.
\end{theorem}

Bernstein's bound is at most as high as Bezout's bound, which is simply
the product of the total degrees,
and is usually significantly smaller
for systems encountered in real-world applications.

The {\em sparse} or {\em Newton resultant} provides a necessary and generically
sufficient condition for the existence of toric roots for a
system of $n+1$ polynomials in $n$ variables:
\begin{equation} \label{inp_sys}
f_1,\ldots,f_{n+1} \in K[x,x^{-1}].
\end{equation}
To define the sparse resultant we
regard a polynomial $f_i$ as a generic point $c_i=(c_{i 1}, \ldots, c_{i m_i})$
in the space of all possible polynomials with the
given support $\A_i=\supp(f_i)$,
where $m_i$ is the number of nonzero terms.
It is natural to identify scalar
multiples, so the space of all such polynomials contracts to the
projective space $\PP_K^{m_i -1}$ or, simply, $\PP^{m_i -1}$.
Then the input system (\ref{inp_sys}) can be thought of as a point
\begin{eqnarray*}
{c} = (c_1,\ldots,c_{n+1})\in \PP^{m_1 -1} \times \cdots \times \PP^{m_{n+1}-1}.
\end{eqnarray*}
Let $Z_0 = Z_0(\A_1, \ldots, \A_{n+1})$ be the set
of all points $c$ such that the system has a solution in
$(\CC^*)^n$ and let $Z = Z(\A_1, \ldots, \A_{n+1})$ denote the
Zariski closure of $Z_0$ in the product of projective spaces.
It is proven in~\cite{PeSt} that $Z$ is an irreducible variety.

\begin{definition}
The {\em sparse resultant} $R=$ $R(\A_1, \ldots, \A_{n+1})$
of system (\ref{inp_sys}) is a polynomial in $\ZZ[{c}]$.
If $\mbox{codim} (Z) = 1$ then $R(\A_1, \ldots, \A_{n+1})$ is the defining
irreducible polynomial of the hypersurface $Z$.
If $\mbox{codim} (Z) > 1$ then $R(\A_1, \ldots, \A_{n+1}) = 1$.
\end{definition}

Let $\deg_{f_i} R$ denote the degree of the resultant $R$
in the coefficients of polynomial $f_i$ and let
$$
\MV_{-i} = \MV(Q_1, \ldots, Q_{i-1}, Q_{i+1}, \ldots, Q_{n+1})
\qquad\mbox{for }i=1,\ldots,n+1.
$$
A consequence of Bernstein's theorem is
\begin{theorem} \label{Tin_resdeg} {\rm\cite{PeSt}}
The sparse resultant is separately homogeneous in the coefficients
$c_i$ of each $f_i$ and its degree in these
coefficients equals the mixed volume of the other $n$ Nwton polytopes
i.e.\ $\deg_{f_i} R = \MV_{-i}$.
\end{theorem}

Canny and Emiris~\cite{CaEm,EmCa:issac,EmCa94}
have proposed the first two efficient algorithms for constructing
{\em resultant matrices}\ i.e., matrices in the coefficients
whose determinant is a nontrivial multiple of the sparse resultant.
The first algorithm relies on a mixed subdivision of the
Minkowski Sum, while the second constructs the matrix in an
incremental fashion.
For those cases where it is provably possible, the incremental
algorithm yields optimal matrices, so that the
determinant equals the resultant.
For general systems, it typically produces matrices that
are at most 3 times larger than optimal.
Let
$$
Q = Q_1+ \cdots + Q_{n+1} \subset \RR^n
$$
be the Minkowski sum of all Newton polytopes.
Let the resultant matrix be $M$.
Now let
$$
\E= (Q+\d) \cap \ZZ^n
$$
be the set that indexes the rows and columns of $M$ in a bijective way,
where $\delta\in\QQ^n$ is an arbitrarily small and {\em sufficiently generic}
vector.
Clearly, $M$ has dimension $|\E|$.
The incremental algorithm also indexes the rows and columns with
monomials, or equivalently, lattice points in $\E$.
This algorithm, though, is different and may select some points more
than once.
In practice, this algorithm produces significantly smaller matrices
but we have no formal result on their dimension.
Irrespective of the algorithm applied,
the resultant matrix $M$ has the following properties.

\begin{theorem} {\rm \cite{EmiPhd}}
Matrix $M$ obtained by either algorithm
is well-defined, square, generically nonsingular and its
determinant is divisible by the
sparse resultant $R$.
\end{theorem}

To be more precise, the rows of $M$ are indexed by a pair
composed of a monomial and an input polynomial.
The entries of the respective row are coefficients of this
polynomial.
The degree of $\det M$ in the coefficients of
$f_i$ for $i=1,\ldots,n+1$ is greater or equal to $\MV_{-i}$.

To solve system~(\ref{square_sys}) we define an overconstrained
system by one of the two ways below.
We apply the resultant matrix construction on the new system
and use the following properties.
Let $\I=\I(f_1,\ldots,f_n)$ be the ideal generated by
polynomials~(\ref{square_sys})
and $V=V(f_1,\ldots,f_n)\in(\oK^*)^n$ their variety,
where $\oK$ is the algebraic closure of field $K$.
Generically, $V$ has {\em dimension zero}.
Then, its coordinate ring $K[x,x^{-1}]/\I$ is an $m$-dimensional
vector space over $K$ by theorem~\ref{Tin_bkk}, where
$$
m=\MV(f_1,\ldots,f_n)=\MV(Q_1,\dots,Q_n).
$$
Using the subdivision-based construction it is easy to
show~\cite{Em94} that generically a monomial basis of $K[x,x^{-1}]/\I$
can be found among the monomials of 
$$
Q_1+\cdots+Q_n\subset\RR^n.
$$
Moreover, resultant matrix $M$ produces the {\em multiplication map}
for any given $f_0$.
This is a matrix, or an endomorphism, that serves in computing
in the coordinate ring and essentially allows computation of
the common roots of $f_1=\cdots=f_n=0$.

\section{Adding a Polynomial} \label{Spo_ures}

The problem addressed here is to find all isolated roots $\alpha\in V$
where $V\subset(\oK^*)^n$ is the zero-dimensional variety
of~(\ref{square_sys}),
with cardinality bounded by $m=\MV(Q_1,\ldots,Q_n)$.
In addition to zero-dimensional, the ideal $\I=\I(f_1,\ldots,f_n)$
is assumed to be {\em radical},
or self-radical i.e., $\I=\sqrt{\I}$,
which is equivalent to saying that all roots in $V$ are distinct.
This requirement is weakened later.

An overconstrained system
is obtained by adding extra polynomial $f_0$ to the given system.
We choose $f_0$ to be linear with coefficients $c_{0j}$ and
constant term equal to indeterminate $u$.
$$
f_0 = u + c_{01} x_1 + \cdots + c_{0n} x_n\;\in K[u][x,x^{-1}].
$$
Coefficients $c_{0j}$, $j=1,\ldots,n$, should define an {\em injective} function
$$
f_0 : V \rightarrow \oK : \alpha \mapsto f_0(\alpha).
$$
There are standard deterministic strategies for selecting $c_{0j}$
so that they ensure the injective property.
In practice, coefficients $c_{0j}$ may be randomly distributed in some
range of integer values of size $S>1$, and
a bad choice for $c_{01},\ldots,c_{0n}$ is one that will result in the same value of
$f_0-u$ at two distinct roots $\alpha$ and $\alpha'$.
Assume that $\alpha$ and $\alpha'$ differ in their $i$-th coordinate for some $i>0$,
then fix all choices of $c_{0j}$ for $j\neq i$; the probability of a bad choice
for $c_{0i}$ is $1/S$, and since there are ${m\choose 2}$ pairs of roots, the
total probability of failure for this scheme is
$$
\mbox{Prob}[\mbox{failure}] \leq {m\choose 2}/S\,:\qquad
c_{0j}\in\{1,\ldots,S\},\, j=1,\ldots,n.
$$
It suffices, therefore, to pick $c_{0j}$ from a sufficiently large range
in order to make the probability of success arbitrarily high.
Moreover, it is clear that any choice of coefficients can be tested
deterministically at the end of the algorithm.

Either algorithm for the resultant matrix may be used to build matrix $M$.
As before, the vanishing of $\det M$ is a necessary condition for the
overconstrained system to have common roots.
For $\alpha\in V$, $u$ is constrained
to a specific value determined by the $c_{0j}$ coefficients.
The construction of $M$ is not affected by this definition of $f_0$.
Let monomial set $\E$ index the columns of $M$ and
partition $M$ so that the lower right square submatrix $M_{22}$ depends on $u$
and has size $r$.
Suppose for now that the upper left square submatrix
$M_{11}$ is {\em nonsingular}.
Clearly $r\geq m$ and equality holds if $M$ is obtained
by the subdivision resultant algorithm.
By the construction of $M$ and for an arbitrary $\alpha\in(\oK^*)^n$
\begin{equation} \label{Epo_uposteval}
\left[ \begin{array}{cc} M_{1 1} & M_{1 2} \\
	M_{2 1} & M_{2 2}(u) \\ \end{array} \right]
\left[ \begin{array}{c} \vdots\\ \alpha^q\\ \vdots\\ \end{array} \right]
= \left[ \begin{array}{c} \vdots\\ \alpha^p f_{i_p}(\alpha)\\ \vdots\\ \end{array} \right]
\; :\qquad q,p\in\E, i_p\in\{0,1,\ldots,n\},
\end{equation}
Now $M'(u)=M_{22}(u) - M_{21} M_{11}^{-1} M_{12}$ where
its diagonal entries are linear polynomials in $u$ and
\begin{equation} \label{Epo_uMprime}
M'(u) \left[ \begin{array}{c} \alpha^{b_1}\\ \vdots\\ \alpha^{b_r}\\ \end{array} \right] =
\left[ \begin{array}{c} \alpha^{p_1}f_0(\alpha,u)\\ \vdots\\ \alpha^{p_r}f_0(\alpha,u)\\
	\end{array} \right],
\end{equation}
where $\B=\{b_1,\ldots,b_r\}$ and $\{p_1,\ldots,p_r\}$ index the columns and rows,
respectively, of $M_{22}$ and thus $M'$.
For a root $\alpha\in V$ and for
$$
u=-\sum_{j=1}^n c_{0j} \alpha_j
$$
the right hand side vector in~(\ref{Epo_uMprime}) is null.
Let $v_{\alpha}' = [\alpha^{b_1}, \ldots, \alpha^{b_r}]$ and write $M'(u)=M'+uI$,
where $M'$ now is numeric and $I$ is the $r\times r$ identity matrix.
Then
$$
(M'+uI) v_{\alpha}' = 0 \Rightarrow
\left[ M' - \left(\sum_j c_{0j} \alpha_{ij}\right) I \right] v_{\alpha}' = 0.
$$
This essentially reduces root-finding to an eigenproblem since,
for every solution of the original system, there is an
eigenvalue and eigenvector of $M$ and hence of $M'$.
Below we study how to compute a candidate solution from every
eigenvalue-eigenvector pair.

If the generated ideal $\I$ is radical then every eigenvalue has
{\em algebraic multiplicity} one with probability greater or equal to
$1-{m\choose 2}/S$.
We can weaken the condition that $\I$ be radical
by requiring only that each eigenvalue has
{\em geometric multiplicity} one.
This equals the dimension of the eigenspace associated
with an eigenvalue.
If there exist eigenvalues of higher geometric multiplicity this
technique fails: then we may use the fact
that specializations of the $u$-resultant yield the root coordinates.
Alternatively
we can define an overconstrained system by hiding a variable as in the next section
and derive the root coordinates one by one.

In what follows we assume that all eigenvalues
have unit geometric multiplicity.
Hence it is guaranteed that among the eigenvectors of $M'$ we shall find
the vectors $v_{\alpha}'$ for $\alpha\in V$.
By construction of $M$~\cite{Em94}
each eigenvector $v_{\alpha}'$ of $M'$ contains the values of
monomials $\B$ at some common root $\alpha\in(\oK^*)^n$.
By~(\ref{Epo_uposteval}) we can define vector $v_{\alpha}$ as follows:
\begin{equation} \label{Epo_fullkvec}
M_{11} v_{\alpha} + M_{12} v_{\alpha}' = 0 \Rightarrow
v_{\alpha} = -M_{11}^{-1} M_{12} v_{\alpha}'.
\end{equation}
The size of $v_{\alpha}$ is $|\E|-r$, indexed by $\E\setminus\B$.
It follows that vectors $v_{\alpha}$ and $v_{\alpha}'$ together contain the values of every
monomial in $\E$ at some root $\alpha$.

\begin{theorem} \label{Tpo_getaffind} {\rm \cite{EmiPhd}}
Assume $\E$ spans $\ZZ^n$.
Then there exists a polynomial-time algorithm that finds a subset
of $n+1$ affinely independent points in $\E$.
Given $v_{\alpha}$, $v_{\alpha}'$ and these points, we can compute the coordinates
of root $\alpha\in V(\I)$.
If all $n+1$ independent points are in $\B$ then $v_{\alpha}'$ suffices.
\end{theorem}

In practice, most of the operations described here are not
implemented with (exact)
rational arithmetic but are instead carried out over floating point numbers
of fixed size.
An important aspect of this computation is numerical error, which we discuss
below in a separate
section and in the particular context of specific applications later.

\begin{theorem} {\rm \cite{EmiPhd}}
Suppose that system~(\ref{square_sys}) generates a zero-dimensional radical ideal,
matrix $M$ has been computed such that $M_{11}$ is nonsingular
and $n+1$ affinely independent points in $\E$ have been computed.
Let $r$ be the size of $M'$,
$\mu$ the maximum number of monomials in any support and
$d$ the maximum polynomial degree in a single variable.
Then all common zeros of the polynomial system are approximated in time
$$
O( |\E|^3 + r n^2 \mu \log d ).
$$
\end{theorem}

It is clear that for most systems the arithmetic complexity is dominated by the
first term, namely the complexity of matrix operations and in particular the
eigendecomposition.
Under reasonable assumptions the complexity becomes~\cite{EmiPhd}
$$
2^{O(n)} m^3,\qquad\mbox{where }m=\MV(f_1,\ldots,f_n).
$$
As expected, the complexity is single exponential in $n$ and polynomial in the
number of roots.

One hypothesis concerned the nonsingularity of $M_{11}$.
When it is singular the resultant matrix is regarded as a linear matrix polynomial
in $u$ and the values and vectors of interest are its singular values and right
kernel vectors.
Finding these for an arbitrary matrix polynomial
is discussed in the next section.

\section{Hiding a Variable} \label{Spo_hide}

An alternative way to obtain an overconstrained system from a well-constrained
one is by {\em hiding} one of the original variables in the coefficient field.
Hiding a variable instead of adding an extra polynomial possesses
the advantage of keeping the number of polynomials constant.
Our experience in solving polynomial systems in robotics and vision suggests
that this usually leads to smaller eigenproblems.
The resultant matrix is regarded as a matrix polynomial in the hidden variable and
finding its singular values and kernel vectors generalizes the $u$-polynomial
construction of the previous section.

Again we suppose the ideal is zero-dimensional and radical.
Formally, given
\begin{equation} \label{Epo_hidinput}
f_0,\ldots,f_{n}\in K[x_1,x_1^{-1},\ldots,x_{n+1},x_{n+1}^{-1}]
\end{equation}
we can view this system as
\begin{equation} \label{Epo_hidover}
f_0,\ldots,f_{n}\in K[x_{n+1}][x_1,x_1^{-1},\ldots,x_n,x_n^{-1}],
\end{equation}
which is a system of $n+1$ Laurent polynomials in variables $x_1,\ldots,x_n$.
Notice that we have multiplied all polynomials by sufficiently high powers of
$x_{n+1}$ in order to avoid dealing with denominators in the {\em hidden
variable} $x_{n+1}$.
This does not affect the system's roots in $(\oK^*)^{n+1}$.

The sparse resultant of this system is a univariate polynomial in $x_{n+1}$.
We show below how this formulation reduces the solution of the original
well-constrained system to an eigenproblem.

\begin{theorem}
Assume that $M$ is a sparse resultant matrix for~(\ref{Epo_hidover}),
with the polynomial coefficients specialized.
Let $\alpha=(\alpha_1,\ldots,\alpha_n)\in(\oK^*)^n$ such that
$(\alpha,\alpha_{n+1})\in(\oK^*)^{n+1}$ is a solution of $f_1=\cdots=f_{n+1}=0$.
Then $M(\alpha_{n+1})$ is singular and column vector
$w=[\alpha^{q_1},\ldots,\alpha^{q_c}]$ lies in the right kernel of $M(\alpha_{n+1})$,
where $\E=\{q_1,\ldots,q_c\}\subset\ZZ^n$ are the exponent vectors indexing
the columns of $M$.
\end{theorem}

\proof
For specialized polynomial coefficients,
$M(\alpha_{n+1})$ is singular by definition.
By construction, right multiplication by a vector of the column
monomials specialized
at a point produces a vector of the values of the row polynomials
at this point.
Let the $i$-th row of $M$ contain the coefficients of $x^{p_j} f_j$, then
$$
M(\alpha_{n+1}) w =
M(\alpha_{n+1}) \left[ \begin{array}{c} \alpha^{q_1}\\ \vdots\\ \alpha^{q_c}\\
	 \end{array} \right] =
\left[ \begin{array}{c} \alpha^{p_1} f_{i_1}(\alpha,\alpha_{n+1})\\ \vdots\\
        \alpha^{p_c} f_{i_c}(\alpha,\alpha_{n+1})\\ \end{array} \right]
= \left[ \begin{array}{c} 0\\ \vdots\\ 0\\ \end{array} \right]
\, :\, i_1,\ldots,i_c\in\{0,1,\ldots,n\}.
$$
\qed

Computationally it is preferable to have to deal with as small a matrix as
possible.
To this end we partition $M$ into four blocks so that the upper left submatrix
$M_{11}$ is square, nonsingular and independent of $x_{n+1}$.
Row and column permutations do not affect the matrix properties so
we apply them to obtain a maximal $M_{11}$.

Gaussian elimination of the leftmost set of columns is now possible and
expressed as matrix multiplication, where $I$ is the identity matrix of
appropriate size:
\begin{equation} \label{Epo_hidgauss}
\left[ \begin{array}{cc} I & 0\\
	-M_{2 1}(x_{n+1}) M_{1 1}^{-1} & I\\ \end{array} \right]
\left[ \begin{array}{cc} M_{1 1} & M_{1 2}(x_{n+1}) \\
	M_{2 1}(x_{n+1}) & M_{2 2}(x_{n+1}) \\ \end{array} \right]
= \left[ \begin{array}{cc} M_{1 1} & M_{1 2}(x_{n+1}) \\
			 0  & M'(x_{n+1})   \\ \end{array} \right],
\end{equation}
where 
$$
M'(x_{n+1}) = M_{22}(x_{n+1}) - M_{21}(x_{n+1}) M_{11}^{-1} M_{12}(x_{n+1}).
$$
Let $\B\subset\E$ index $M'$.
We do not have {\it a priori} knowledge
of the sizes of $M_{11}$ and $M'$ whether the subdivision or the incremental
algorithm has constructed $M$.

\begin{corollary}
Let $\alpha=(\alpha_1,\ldots,\alpha_n)\in(\oK^*)^n$ such that $(\alpha,\alpha_{n+1})\in(\oK^*)^{n+1}$ is a
common zero of $f_0=\cdots=f_{n}=0$.
Then $\det M'(\alpha_{n+1}) = 0$ and, for any vector $v'=[\cdots\alpha^{q}\cdots]$,
where $q$ ranges over $\B$,
$M'(\alpha_{n+1}) v'=0$.
\end{corollary}

To recover the root coordinates, $\B$ must affinely span $\ZZ^n$, otherwise
we have to compute the kernel vector of matrix $M$ which equals the
concatenation of vectors $v$ and $v'$, where $v'$ is the kernel vector
of $M'$ and $v$ is specified from~(\ref{Epo_hidgauss}):
\begin{eqnarray*}
\left[ \begin{array}{cc} M_{1 1} & M_{1 2}(\alpha_{n+1}) \\
			 0  & M'(\alpha_{n+1})   \\ \end{array} \right]
\left[ \begin{array}{c} v\\ v'\\ \end{array} \right]
= \left[ \begin{array}{c} 0\\ 0\\ \end{array} \right] & \Rightarrow &
M_{11} v + M_{12}(\alpha_{n+1}) v' = 0\\
& \Leftrightarrow &
v = - M_{11}^{-1} M_{12}(\alpha_{n+1}) v',
\end{eqnarray*}
since $M_{11}$ is defined to be the maximal nonsingular submatrix.
The concatenation $[v,v']$ is indexed by $\E$
which always includes an affinely independent subset unless all
$n$-fold mixed volumes are zero and no roots exist.
Then, we recover all root coordinates
by taking ratios of the vector entries.

We now concentrate on matrix polynomials and their companion matrices;
for definitions and basic results consult~\cite{GLR}.
Denote the hidden variable by $x$, then
\begin{equation} \label{Epo_hidx}
f_i\in K[x][x_1,x_1^{-1},\ldots,x_n,x_n^{-1}],\qquad i=0,\ldots,n,
\end{equation}
and we denote the univariate matrix $M'(x_{n+1})$ by $A(x)$.
Let $r$ be the size of $A$,
and $d\geq 1$ the highest degree of $x$ in any entry.
We wish to find all values for $x$ at which matrix
$$
A(x)=x^d A_d + x^{d-1} A_{d-1} + \cdots + x A_1 + A_0
$$
becomes singular, where matrices $A_d,\ldots,A_0$ are all square, of order
$r$ and have numerical entries.
We refer to $A(x)$ as a {\em matrix polynomial} with degree $d$ and
matrix coefficients $A_i$.
The values of $x$ that make $A(x)$ singular are its {\em eigenvalues}.
For every eigenvalue $\l$, there is a basis of the kernel of $A(\l)$ defined by
the {\em right eigenvectors} of the matrix polynomial associated to $\l$.
This is the eigenproblem for matrix polynomials, a classic problem in linear
algebra.

If $A_d$ is nonsingular then the eigenvalues and right eigenvectors of $A(x)$
are the eigenvalues and right eigenvectors of a {\em monic} matrix polynomial.
Notice that this is always the case with the $u$-resultant formulation in the
previous section.
$$
A_d^{-1} A(x) = x^d I + x^{d-1} A_d^{-1} A_{d-1} + \cdots +
	x A_d^{-1} A_1 + A_d^{-1} A_0,
$$
where $I$ is the $m\times m$ identity matrix.
The {\em companion matrix} of this monic matrix polynomial is defined to be
a square matrix $C$ of order $rd$.
$$
C = \left[ \begin{array}{cccc}
	0       & I     & \cdots        & 0     \\
	\vdots  &       & \ddots        &       \\
	0       & 0     & \cdots        & I     \\
	-A_d^{-1}A_0    & -A_d^{-1}A_1 & \cdots & -A_d^{-1}A_{d-1} \\
	\end{array} \right].
$$
It is known that
the eigenvalues of $C$ are precisely the eigenvalues of the monic polynomial,
whereas its right eigenvectors contain as subvectors the right eigenvectors
of $A_d^{-1} A(x)$.
Formally, assume
$w=[v_1,\ldots,v_d]\in\oK^{md}$ is a (nonzero) right eigenvector of $C$,
where each $v_i\in\oK^m$, $i=1,\ldots,d$.
Then $v_1$ is a (nonzero) right eigenvector of $A_d^{-1} A(x)$ and
$v_i=\l^{i-1} v_1$, for $i=2,\ldots,d$, where $\l$ is the eigenvalue of $C$
corresponding to $w$.

We now address the question of a singular leading matrix in a non-monic polynomial.
The following transformation is also used in the implementation in order to improve the
conditioning of the leading matrix.

\begin{lemma} \label{Lpo_transform}
Assume matrix polynomial $A(x)$ is not identically singular for all $x$
and let $d$ be the highest degree in $x$ of any entry.
Then there exists a transformation
$x\mapsto (t_1 y + t_2)/(t_3 y + t_4)$ for some
$t_1,t_2,t_3,t_4\in\ZZ$, that produces a new matrix polynomial $B(y)$
of the same degree and with matrices of the same dimension, such that
$B(y)$ has a nonsingular leading coefficient matrix.
\end{lemma}

It is easy to see that the resulting polynomial $B(y)$ has matrix coefficients
of the same rank, for sufficiently generic scalars $t_1,t_2,t_3,t_4$,
since every matrix is the sum of $d+1$ scalar products of $A_i$.
Thus this transformation is often referred to as {\em rank balancing}.

\begin{theorem}
If the values of the hidden variable in the solutions of
$f_1=\cdots=f_{n+1}=0$, which correspond to eigenvalues of the matrix polynomial,
are associated to eigenspaces of unit dimension
and $A(x)$ is not identically singular then we
can reduce root-finding to an eigenproblem and some evaluations of the input
polynomials at candidate roots.
\end{theorem}

\proof
We have seen that the new matrix polynomial $B(y)$
has a nonsingular leading coefficient.
Moreover, finding its eigenvalues and eigenvectors is reduced
to an eigenproblem of the companion matrix $C$.
By hypothesis,
the eigenspaces of $C$ are one-dimensional therefore for every root
there is an eigenvector that
yields $n$ coordinates of the root by theorem~\ref{Tpo_getaffind}.
The associated eigenvalue may be either simple or multiple and yields
the value of the hidden variable at the same root.
Notice that extraneous eigenvectors and eigenvalues may have to be rejected by
direct evaluation of the input polynomials at the candidate roots
and a zero test.
The right eigenvectors of $B(y)$ are identical to those of $A(x)$ but any
eigenvalue $\l$ of the former
yields $(t_1\l +t_2)/(t_3\l +t_4)$ as an eigenvalue of $A(x)$.
\qed

The condition that all eigenspaces are unit-dimensional is equivalent to
the solution coordinate at the hidden variable having unit geometric multiplicity.
For this it suffices that the algebraic multiplicity of these solutions be
one i.e., all hidden coordinates must be distinct.
Since there is no restriction in picking which variable to hide, it is enough
that one out of the original $n+1$ variables have unit geometric multiplicity.
If none can be found, we can specialize the hidden variable to each of the
eigenvalues and solve every one of the resulting subsystems.

Our complexity bounds shall occasionally ignore the logarithmic terms; this is
expressed by the use of $O^*(\cdot)$.
Let $\mu$ the maximum number of monomials in any ($n$-variate) polynomial
of~(\ref{Epo_hidover}) and
$d$ the maximum polynomial degree in a single variable; this is typically
larger than the highest degree of the hidden variable but in the worst case
they are equal.

\begin{theorem} {\rm \cite{EmiPhd}}
Suppose that the ideal of~(\ref{Epo_hidover}) is zero-dimensional,
the coordinates of the hidden variable are distinct,
matrix $M$ has been computed such that $M_{11}$ is nonsingular and the resulting
matrix polynomial $A(x)$ is regular.
In addition, a set of affinely independent points in $|\E|$ has been computed.
Then all common isolated zeros are computed with worst-case complexity
$$
O^*( |\E|^3 d + \MM(rd) + rd n^2 \mu ) =
O^*( |\E|^3 d^3 + |\E| d n^2 \mu ).
$$
\end{theorem}

Under some reasonable assumptions about the input
the complexity becomes~\cite{EmiPhd}
$$
2^{O(n)} O(m^6),\qquad\mbox{where }m=\MV(f_1,\ldots,f_n).
$$
It is clear by the first bound on arithmetic complexity that the most expensive
part is the eigendecomposition of the companion matrix.
Moreover, the problem of root-finding is exponential in $n$ as expected.

\section{Implementation and Numerical Accuracy} \label{Spo_numerical}

Our implementation is entirely in Ansi~C.
The overall method has two stages, one online and one offline.
A program of independent interest, which makes part of this
package, is already available.
It is the implementation of 
a fast algorithm for computing mixed volumes~\cite{EmCa94}
and can be obtained by anonymous {\tt ftp} on
{\tt robotics.eecs.Berkeley.edu}, from directory {\tt MixedVolume}.

One advantage of the resultant method
over previous algebraic as well as numerical methods is that
the resultant matrix need only be
computed once for all systems with the same set of exponents.
So this step can often be done offline, while the eigenvalue calculations to
solve the system for each coefficient specialization are online.

Another offline operation is to permute rows and columns of the given resultant
matrix in order to create a maximal square submatrix at the upper left corner
which is independent of the hidden variable.
In order to minimize the computation involving polynomial entries and to reduce
the size of the upper left submatrix that must be inverted, the program
concentrates all constant columns to the left and within these columns
permutes all zero rows to the bottom.

To find the eigenvector entries that will allow us to recover the
root coordinates it is typically sufficient to examine $\B$ indexing $M_{22}$
and search for pairs of entries corresponding to exponent vectors
$v_1, v_2$ such that $v_1-v_2=(0.\ldots,0,1,0,\ldots,0)$.
This will let us compute the $i$-th coordinate if the unit appears
at the $i$-th position.
For inputs where we need to use points in $\E\setminus\B$, only
the entries indexed by these points must be computed in vector
$v_{\alpha}$, in the notation of theorem~\ref{Tpo_getaffind}.
In any case the complexity of this step is dominated.

The input to the online solver is the set of all supports, the associated coefficients and
matrix $M$ whose rows and columns are indexed by monomial sets.
The output is a list of candidate roots and the values of each at the given
polynomials.
The test for singularity of the matrix polynomial is implemented by substituting
a few random values in the hidden variable and testing whether the resulting
numeric matrix has full rank.
For rational coefficients this is done in modular arithmetic so the answer
is exact within a finite field, therefore nonsingularity is decided with
very high probability.

Most of the online computation is numeric for reasons of speed;
in particular we use double precision floating point numbers.
We use the LAPACK package~\cite{LAPACK} because it implements state-of-the-art
algorithms, including block algorithms on appropriate architectures,
and provides efficient ways for computing condition numbers and
error bounds.
Moreover it is publicly available, portable, includes a C version and is soon
to include a parallel version for shared-memory machines.
Of course it is always possible to use other packages
such as EISPACK~\cite{eispack} or LINPACK~\cite{linpack}.

A crucial question in numerical computation is to predict the extent of
roundoff error; see for instance~\cite{GoVL}.
To measure this we make use of the standard definition of matrix norm
$\| A \|_p$ for matrix $A$.
We define four condition numbers for $A$:
$$
\kappa_p(A) = \| A \|_p \| A^{-1} \|_p,\;\;
p=1,2,\infty,F,\;\; A \mbox{ square},\qquad\mbox{and if } A
\mbox{ singular}\,\; \kappa_p(A) = \infty,
$$
where $F$ denotes the Frobenius norm.
$\kappa_2(A)$ equals the ratio of the maximum over the minimum 
singular value of $A$.
For a given matrix, the ratio of any two condition numbers with $p=1,2,\infty,F$
is bounded above and below by the square of the matrix dimension or its inverse.
We also use $\|v\|_p$ for the $p$-norm of vector $v$.

As precise condition numbers are sometimes expensive to compute, LAPACK provides
approximations of them and the associated error bounds.
Computing these estimates is very efficient compared to the principal computation.
These approximations run the risk of underestimating the correct values,
though this happens extremely seldom in practice.
Error bounds are typically a function of the condition number and the matrix
dimension; a reasonable assumption for the dependence on the latter is to
use a linear function.
Small condition numbers indicate well-behaved or {\em well-conditioned} matrices,
e.g.\ orthogonal matrices are perfectly conditioned with $\kappa=1$.
Large condition numbers characterize {\em ill-conditioned} matrices.

After permuting rows and columns so that the maximal upper left submatrix
is independent of $u$ or the hidden variable $x$,
we apply an LU decomposition with column pivoting to the upper left submatrix.
We wish to decompose the maximal possible submatrix so that it has a
reasonable condition number.
The maximum magnitude of an acceptable condition number can be controlled by the
user; dynamically, the decomposition stops when the pivot takes a value smaller
than some threshold.

To compute $M'$ we do not explicitly form $M_{11}^{-1}$ but use its
decomposition to solve linear problem $M_{11} X = M_{12}$ and then
compute $M' = M_{22} - M_{21} X$.
Different routines are used, depending on $\kappa(M_{11})$, to solve the
linear problem, namely the slower but more accurate {\tt dgesvx} function
is called when this condition number is beyond some threshold.
Let $\hx_{j}$ denote some column of $X$ and let $x_{j}$ be the respective
column if no roundoff error were present.
Then the error is bounded~\cite{GoVL} by
$$
\frac{ \| x - \hx \|_{\infty} }{ \| x \|_{\infty} }
\lid 4 \cdot 10^{-15} (|\E|-r) \kappa_{\infty}(M_{11}),
$$
where $|\E|-r$ is the size of $M_{11}$ and we have used
$2\cdot 10^{-16}$ as the machine precision under the IEEE double precision
floating point arithmetic.

For nonsingular matrix polynomials $A(x)$ we try a few random integer quadruples 
$(t_1,t_2,t_3,t_4)$.
We then redefine the matrix polynomial to be
the one with lowest $\kappa(A_d)$.
This operation of {\em rank balancing} is indispensable when
the leading coefficient is nonsingular as well as ill-conditioned.
Empirically we have observed that for matrices of dimension larger than
200, at least two or three quadruples should be tried since a
lower condition number by two or three orders of magnitude is
sometimes achieved.
The asymptotic as well as practical
complexity of this stage is dominated by the other stages.

If we manage to find a matrix polynomial with well-conditioned $A_d$
we compute the equivalent monic polynomial and call the standard
eigendecomposition routine.
There are again two choices in LAPACK for solving this problem with an
iterative or a direct algorithm, respectively implemented in routines
{\tt hsein} and {\tt trevc}.
Experimental evidence points to the former as being faster on 
problems where the matrix size is at least 10 times larger than
the mixed volume, since an iterative solver can better exploit the fact
that we are only interested in real eigenvalues and eigenvectors.

If $A_d$ is ill-conditioned for all linear $t_i$ transformations
we build the matrix pencil and call the
{\em generalized eigendecomposition} routine {\tt dgegv} to solve $C_1x + C_0$.
The latter returns pairs $(\alpha,\beta)$ such that matrix
$C_1 \alpha + C_0 \beta$ is singular.
For every $(\alpha,\beta)$ pair there is a nonzero right generalized eigenvector.
For nonzero $\beta$ we obtain the eigenvalues as $\alpha/\beta$, while for zero
$\beta$ and nonzero $\alpha$ the eigenvalue tends to infinity and depending on
the problem at hand we may or may not wish to discard it.
The case $\alpha=\beta=0$ occurs if and only if the pencil is identically zero
within machine precision.

In recovering the eigenvector of matrix polynomial $A(x)$ from the eigenvector
of its companion matrix, we can use any subvector of the latter.
We choose the topmost subvector when the eigenvalue is smaller than the
unit, otherwise we use one of the lower subvectors.
Nothing changes in the algorithm, since ratios of the entries will still
yield the root, yet the stability of these ratios is improved.

There are certain properties of the problem that have not been exploited yet.
Typically, we are interested only in real solutions.
We could concentrate, therefore, on the real eigenvalues and eigenvectors
and choose the algorithms that can distinguish between them and complex
solutions at the earliest stage.
Moreover, bounds on the root magnitude may lead to substantial savings, as
exemplified in~\cite{Ma94}.
In the rest of this article we examine these issues in the light of concrete
applications of our program.

\section{Camera Motion from Point Matches} \label{Sapp_pts2motion}

{\em Camera motion reconstruction}, or {\em relative orientation},
in its various forms is a basic problem in photogrammetry, including
the computation of the shape of an object from its motion.
Formally, we are interested in the problem of computing the displacement of a
camera between two positions in a static environment.
Given are the coordinates of
certain points in the two views under perspective projection
on calibrated cameras.
Equivalently, this problem consists in computing the displacement of a
rigid body, whose identifiable features include only points,
between two snapshots taken by a stationary camera.
We consider, in particular, the case where the minimum number of
5 point matches is available.
In this case the algebraic problem reduces to a well-constrained system of
polynomial equations and we are able to give a closed-form solution.

Typically, computer vision applications use at least 8 points in order to
reduce the number of possible solutions to 3 and, for generic configurations,
to one.
In addition, computing the displacement reduces to a linear problem and the
effects of noise in the input can be diminished~\cite{L-H81}.
Our approach shows performance comparable to these methods and, as it requires
the minimum number of data points, it
is well-suited for detecting and eliminating {\em outliers} in the presence of noise.
These are data points which are so much affected by noise that they should
not be taken into account.

\subsection{Algebraic Formulation}	\label{Svis_quat}

To formalize, let orthonormal $3\times 3$ matrix
$R\in\SO(3,\RR)$ denote the rotation.
Let (column) vector $t\in\RR^3$ denote
the camera translation in the original frame of reference.
The 5 points in the two images are (column) vectors $a_i,a_i'\in\PP_{\RR}^2$,
for $i=1,\ldots,5$
in the first and second frame of reference respectively.
It is clear that the magnitude of $t$ cannot be recovered, hence there are
5 unknowns, 3 defining the rotation and 2 for the translation.

The following quaternion formulation was independently suggested by
J.~Canny and~\cite{Horn91}.
Let $x,y,z$ be 3-vectors and
$\qx=[x_0,x],\qy=[y_0,y],\qz=[z_0,z]\in\RR^4$ be arbitrary quaternions.
Let $\qx\qy$ represent a quaternion product
and $\qz^*=[z_0,-z]$ be the conjugate quaternion of $\qz$.
Quaternions $\qq=[q_0,q],\,\qt=[t_0,t]\in\RR^4$ represent the
rotation and translation respectively.
A rotation represented by angle $\phi$ and unit 3-vector $s$ is expressed
uniquely by quaternion $[\cos \phi/2, \sin \phi/2 \, s]$, hence
any rotation quaternion has unit 2-norm.

\begin{table}
\begin{center}
\begin{tabular}{|c|c|c|}
\hline
system & operation & CPU time \\
\hline \hline
$6\times 6$ & mixed volume                      & 1m 16s\\
\hline \hline

$6\times 5$ & sparse resultant (offline)        & 12s\\
\hline \hline

$6\times 6$ (first) & root finding (online)             & 0.2s\\
\hline
$6\times 6$ (second) & root finding (online)             & 1s (\sparc)\\
\hline
\end{tabular}
\caption{Camera motion from point matches: running times are measured
	on a \deca\ except for the second system which is solved on a \sparc.
	\label{Tab_motion}}
\end{center} \end{table}

The equations in terms of the quaternions are homogeneous.
After dehomogenization
we obtain 6 polynomials in
6 variables organized in two 3-vectors.
We denote these vectors by $q,d$.
\begin{eqnarray}
(a_i^T q) (d^T a_i') + a_i^T a_i' +
(a_i\times q)^T a_i' + (a_i\times q)^T (d\times a_i') +
a_i^T (d\times a_i') & = & 0,\;\; i=1,\ldots,5 \nonumber \\
1 - d^T  q & = & 0              \label{Emot_sys2}
\end{eqnarray}

This system is well-constrained so we can apply Bernstein's bound to approximate
the number of roots in $(\CC^*)^6$.
The mixed volume of this system is 20, which is an exact bound~\cite{Dem88}.
The performance of our implementation on mixed volume
is shown in table~\ref{Tab_motion} for the \deca\
of table~\ref{Tin_hard}.

\subsection{Applying the Resultant Solver}

Resultant matrix $M$ is of dimension $60$, while only 20 columns contain
hidden variable $q_1$.
$40\times 40$ submatrix $M_{11}$ is inverted
and the resulting $20\times 20$ pencil
is passed to an eigendecomposition routine.
The running times for the two main phases are reported in table~\ref{Tab_motion}.
The rest of this section examines the accuracy of our procedure
on two specific instances from~\cite{FaMa90}.
We use error criterion
\begin{equation} \label{Emot_errcrit}
\sum_{i=1}^5 \frac{1}{ \| a_i \| \cdot \| a_i' \| }
| a_i^T (t\times R a_i') |,
\end{equation}
where $|\cdot|$ denotes absolute value.
$\| \cdot \|$ is the vector 2-norm
and $t$, $R$ are the calculated translation and rotation.
This expression vanishes when the solutions to $R,t$ are exact,
otherwise it returns the absolute value of error normalized by
the input norm.

\begin{table}
\begin{center} \begin{tabular}{|c|r|r|r|r|}
\hline
machine & clock rate [MHz] & memory [MB] & SpecInt92 & SpecFP92\\
\hline \hline
{\sc DEC Alpha 3000/300} & 150 & 64 & 67 & 77\\
\hline
{\sc Sun Sparc 20/61}    & 60 & 32 & 95 & 93\\
%% \hline
%% {\sc Sun Sparc 10/40}    & 40 & 32 & 50 & 60\\
\hline
\end{tabular}
\caption{Hardware specifications}       \label{Tin_hard}
\end{center} \end{table}

The first example admits the maximum of 10 pairs of real solutions.
$M_{11}$ is well-conditioned 
with $\kappa < 10^3$, so it is factorized to yield an optimal pencil
of dimension 20.
The latter has a well-conditioned leading coefficient with $\kappa < 10^4$,
so it yields a monic polynomial
on which the standard eigendecomposition is applied.
The maximum value of error criterion~(\ref{Emot_errcrit}) is less than
$1.3\cdot 10^{-6}$, which is very satisfactory.

The input parameters are sufficiently generic in order to lead to the maximum
number of real solutions.
This explains the good conditioning of the matrices.
The second example looks at a less generic instance: namely
a configuration that has only 8 distinct real solutions
which are clustered close together.
This is manifest in the fact that matrix $M_{11}$ is ill-conditioned. 
The input is constructed by applying a rotation
of $60.00145558\degs$ about the $z$ axis in the first frame
and a translation of $(.01,.01,-1)$.

Exactly the same procedure is used as before
to produce a $60\times 60$ resultant matrix,
but now the upper leftmost $40\times 40$ submatrix $M_{11}$ has $\kappa >.6 \cdot 10^5$
so we choose not to invert it.
Instead, the $60\times 60$ pencil is considered: the leading matrix has
$\kappa > 10^8$ in its original form and for any of 4 random transformations,
hence it is not inverted and the generalized eigendecomposition is applied.

Applying error criterion~(\ref{Emot_errcrit}), the largest
absolute value is $7.4\cdot 10^{-5}$, hence all 16 real solutions are quite accurate.
There are another 4 complex solutions and 40 infinite ones.
The total CPU running time for the online phase is, on the average,
1 second on the \sparc\ of table~\ref{Tin_hard}.

It is interesting to observe 
in connection to roots with zero coordinates, that
in the second example our solver recovers roots in
$\CC^n \setminus (\CC^*)^n$,
namely we find a camera motion whose rotation quaternion $\qq$
has $q_1=q_2=0$.
Such roots can be thought of as limits of roots in $(\CC^*)^n$ as
the system coefficients deform.
As long as the variety does not generically reside in
$\CC^n \setminus (\CC^*)^n$,
roots with zero coordinates will always be recovered.
This is typically the case in practical applications.
For the particular example, there is a stronger reason why all roots
are recovered, namely all polynomials include a constant term
(see theorem~\ref{Tallroots}).

There exist various implementations of linear methods requiring at least
8 points, including Luong's~\cite{Luo92}.
We have been able to experiment with this program which implements the
least-squares method of~\cite{TH84}, and found it faster on both instances
but less accurate on the second one.
In particular, we chose specific solutions
to generate an additional 3 matches for each of the problems above.
The average CPU time on the two examples is
$0.08$ seconds on a \sparc.
On the first example
the output is accurate to at least 7 digits.
On the second example, Luong's implementation returns a rotation
of $60.0013\degs$ about $(-10^{-5}, 10^{-5}, 1)$ and a unit translation
vector of $(-10^{-5}, -10^{-5}, -1)$.
But the latter differs significantly from the true vector
$t=(.01,.01,-1)$.

\section{Conformational Analysis of Cyclic Molecules} \label{Sapp_mole}

A relatively new branch of computational biology has been emerging as an effort
to apply successful paradigms and techniques from geometry and
robot kinematics to predicting the structure
of molecules, embedding them in euclidean space and finding the energetically
favorable configurations~\cite{PC:94,EmiPhd}.
The main premise for this interaction is the observation that
various structural requirements on molecules
can be modeled as geometric or kinematic constraints.

This section examines the problem of computing all {\em conformations} of a
{\em cyclic} molecule, which reduces to an {\em inverse kinematics} problem.
Conformations specify the 3-dimensional structure of the molecule.
It has been argued by G\=o and Scherga~\cite{GoSc70} that energy minima can
be approximated by allowing only the dihedral angles to vary, while keeping
bond lengths and bond angles fixed.
At a first level of approximation, therefore, solving for the dihedral angles
under the assumption of {\em rigid geometry}
provides information for the energetically favorable configurations.

We consider molecules of six atoms to illustrate our approach
and show that the corresponding algebraic formulation conforms to our model
of sparseness.
Our resultant solver is able to compute all solutions accurately
even in cases where multiple solutions exist.

\subsection{Algebraic Formulation}
\label{Sapp_molform}

The molecule has a cyclic backbone of 6 atoms, typically of carbon.
They determine primary structure, the object of our study.
Carbon-hydrogen or other bonds outside the backbone are ignored.
The bond lengths and angles provide the constraints while the six dihedral angles
are allowed to vary.
In kinematic terms, atoms and bonds are analogous to {\em links} and
{\em joints} of a {\em serial mechanism} in which each pair of consecutive
axes intersects at a link.
This implies that the link offsets are zero for all six links
which allows us to reduce the 6-dimensional problem to a system of
3 polynomials in 3 unknowns.
The product of all link transformation matrices is the
identity matrix, since the end-effector is at the same position and orientation
as the base link.

\begin{figure}
\centerline{\psfig{figure=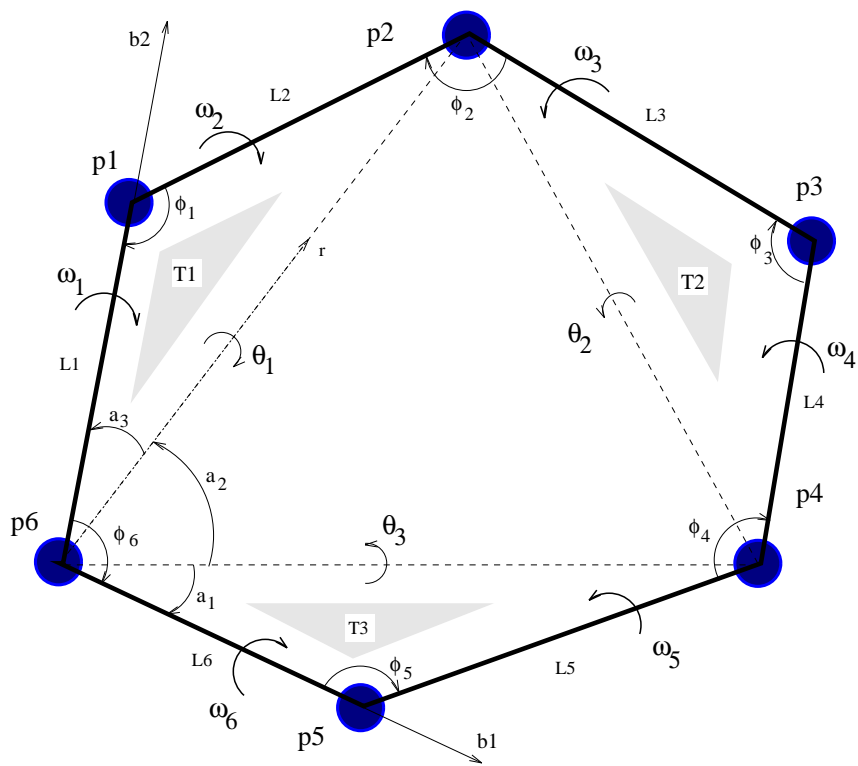,height=3.95in} }
\caption{The cyclic molecule.\label{Fring_def}}
\end{figure}

We adopt an approach proposed by D.~Parsons~\cite{Par94}.
Notation is defined in figure~\ref{Fring_def}.
Backbone atoms are regarded as points $p_1,\ldots,p_6 \in\RR^3$;
the unknown dihedrals are the angles $\omega_1,\ldots,\omega_6$
about axes $(p_6,p_1)$ and $(p_{i-1},p_i)$ for $i=2,\ldots,6$.
For readers familiar with the kinematics terminology, the
Denavit-Hartenberg parameters are
$$
\alpha_i = 180\degs-\phi_i,
d_i = L_i,
a_i = 0,
\theta_i = \omega_i.
$$
Each of triangles $T_1=\triangle(p_1,p_2,p_6)$, $T_2=\triangle(p_2,p_3,p_4)$ and
$T_3=\triangle(p_4,p_5,p_6)$ is fixed
for constant bond lengths $L_1,\ldots,L_6$ and bond angles $\phi_1,\phi_3,\phi_5$.
Then the lengths of $(p_2,p_6)$, $(p_2,p_4)$ and $(p_4,p_6)$ are constant,
hence {\em base} triangle $\triangle(p_2,p_4,p_6)$ is fixed in space,
defining the $xy$-plane of a coordinate frame.
Let $\theta_1$ be the (dihedral) angle between the plane of $\triangle(p_1,p_2,p_6)$
and the $xy$-plane.
Clearly, for any conformation $\theta_1$ is well-defined.
Similarly we define angles $\theta_2$ and $\theta_3$, as shown in figure~\ref{Fring_def}.
We call them {\em flap} (dihedral) angles to distinguish them from the bond dihedrals.

Conversely, given lengths $L_i$, angles $\phi_i$ for $i=1,\ldots,6$ and
flap angles $\theta_i$ for $i=1,\ldots,3$ the coordinates of all points $p_i$ are
uniquely determined and hence the bond dihedral angles and the associated
conformation are all well-defined.
We have therefore reduced the problem to computing the three flap angles $\theta_i$
which satisfy the constraints on bond angles $\phi_2, \phi_4, \phi_6$.

Hence we obtain polynomial system
\begin{eqnarray}        \label{Ering_6system}
\alpha_{11} + \alpha_{12} \cos \theta_2 + \alpha_{13} \cos \theta_3 +
	\alpha_{14} \cos \theta_2 \cos \theta_3 +
	\alpha_{15} \sin \theta_2 \sin \theta_3 & = & 0, \nonumber \\
\alpha_{21} + \alpha_{22} \cos \theta_3 + \alpha_{23} \cos \theta_1 +
	\alpha_{24} \cos \theta_3 \cos \theta_1 +
	\alpha_{25} \sin \theta_3 \sin \theta_1 & = & 0, \nonumber \\
\alpha_{31} + \alpha_{32} \cos \theta_1 + \alpha_{33} \cos \theta_2 +
	\alpha_{34} \cos \theta_1 \cos \theta_2 +
	\alpha_{35} \sin \theta_1 \sin \theta_2 & = & 0,\\
\cos^2 \theta_1 + \sin^2 \theta_1 - 1 & = & 0, \nonumber       \\
\cos^2 \theta_2 + \sin^2 \theta_2 - 1 & = & 0, \nonumber       \\
\cos^2 \theta_3 + \sin^2 \theta_3 - 1 & = & 0, \nonumber
\end{eqnarray}
where the $\alpha_{ij}$ are input coefficients.

This system has Bezout bound of 64 and mixed volume 16;
the mixed volume is the exact number of complex roots generically
as we shall prove below by demonstrating an instance with 16 real roots.
Notice that 16 is also the exact number of solutions, generically, to the
{\em general} inverse kinematics problem with 6 rotational joints (6R).

For our resultant solver we prefer an equivalent formulation with a
smaller number of polynomials, obtained by applying the standard
transformation to half-angles that gives {\em rational} equations in the new
unknowns $t_i$:
$$
t_i = \tan\frac{\theta_i}{2}\; :\;\;\;\;\cos \theta_i = \frac{1-t_i^2}{1+t_i^2},\;
\sin \theta_i = \frac{2t_i}{1+t_i^2},\qquad i=1,2,3.
$$
This transformation captures automatically the last three equations
in~(\ref{Ering_6system}).
By multiplying both sides of the $i$-th equation by $(1+t_j^2)(1+t_k^2)$,
where $(i,j,k)$ is a permutation in $S(1,2,3)$, the polynomial system becomes
\begin{eqnarray}        \label{Ering_3system}
f_1 = \beta_{11} + \beta_{12} t_2^2 + \beta_{13} t_3^2 +
	\beta_{14} t_2^2 t_3^2 + \beta_{15} t_2 t_3 & = & 0 \nonumber \\
f_2 = \beta_{21} + \beta_{22} t_3^2 + \beta_{23} t_1^2 +
	\beta_{24} t_3^2 t_1^2 + \beta_{25} t_3 t_1 & = & 0 \\
f_3 = \beta_{31} + \beta_{32} t_1^2 + \beta_{33} t_2^2 +
	\beta_{34} t_1^2 t_2^2 + \beta_{35} t_1 t_2 & = & 0 \nonumber
\end{eqnarray}
where $\beta_{ij}$ are input coefficients.
The new system has again Bezout bound of 64 and mixed volume 16.

\subsection{Applying the Resultant Solver} \label{Sapp_molsolve}

The first instance is a synthetic example for which we fix one feasible
conformation with all flap angles equal to $90\degs$.
All polynomials are multiplied by 8 in order for the coefficients
to be all integers, then $\beta_{ij}$ is the $(i,j)$-th entry of matrix
$$
\left[ \begin{array}{rrrrr}
	-9 & -1 & -1 & 3 & 8 \\
	-9 & -1 & -1 & 3 & 8 \\
	-9 & -1 & -1 & 3 & 8 \\
\end{array} \right].
$$
The symmetry of the problem is bound to produce root
coordinates of high multiplicity, so we decide to follow the first approach
to solving the system (sect.~\ref{Spo_ures}) and add polynomial
$$
f_0 = u + 31 t_1 - 41 t_2 + 61 t_3
$$
with randomly selected coefficients.
In this system, the 3-fold mixed volumes are $12,12,12,16$ hence
the sparse resultant has total degree 52 and degree 16 in $f_0$.
The resultant matrix is regular and
has dimension 86, with 30 rows corresponding to $f_0$.
This is the offline phase; the rest corresponds to the online
execution of the solver.

The entire $56\times 56$ upper left submatrix is decomposed and is relatively
well-conditioned.
In the $30\times 30$ matrix polynomial,
the leading matrix coefficient is singular within machine precision;
two random transformations are used but fail to improve significantly
the conditioning of the matrix.
Therefore the
generalized eigenproblem routine is called on the $30\times 30$
pencil and produces 12 complex solutions, 3 infinite real solutions
and 15 finite real roots.
The absolute value of the four polynomials on the
candidate values lies in $[0.6\cdot 10^{-9}, 0.3\cdot 10^{-3}]$
for values that approximate true solutions and in
$[7.0,3.0\cdot 10^{20}]$ for spurious answers.
Our program computes the true roots to at least 5 digits as
seen by comparing with the exact solutions computed by
\MapleV\ using Gr\"obner bases over the rationals.
The latter are
$$
\pm(1,1,1), \pm(5,-1,-1), \pm(-1,5,-1), \pm(-1,-1,5).
$$
The average CPU time of the online phase on the \sparc\ of table~\ref{Tin_hard}
is $0.4$ seconds.

Usually noise enters in the process that produces the coefficients;
this example models this phenomenon.
We consider the {\em cyclohexane} molecule which has 6 carbon atoms
at equal distances and equal bond angles.
Starting with the pure cyclohexane,
we randomly perturb them by about $10\%$
to obtain $\beta_{ij}$ as the entries of matrix
$$
\left[ \begin{array}{rrrrr}
	-310& 959 & 774& 1313& 1389\\
	-365& 755 & 917& 1269& 1451\\
	-413& 837 & 838& 1352& 1655
\end{array} \right].
$$
We used the second approach to define an overconstrained system, 
namely by hiding variable $t_3$ in the coefficient field (sect.~\ref{Spo_hide}).
The resultant matrix has
dimension 16 and is quadratic in $t_3$,
whereas the 2-fold mixed volumes are all 4 and the sparse resultant has degree
$4+4+4=12$.

The monic quadratic polynomial reduces to a $32\times 32$ companion matrix
on which the standard eigendecomposition is applied.
After rejecting false candidates, the recovered roots
cause the maximum absolute
value of the input polynomials to be $10^{-5}$.
We check the computed solutions against those obtained by a
Gr\"obner bases computation over the integers
and observe that each contains at least 8 correct digits.
The total CPU time on a \sparc\ is $0.2$ seconds on average for the
online phase.

Lastly we report on an instance where the input parameters are
sufficiently generic to produce 16 real roots.
The $\beta_{ij}$ coefficients are given by matrix
$$
\left[ \begin{array}{rrrrr}
	-13 & -1 & -1 & -1 & 24\\
	-13 & -1 & -1 & -1 & 24\\
	-13 & -1 & -1 & -1 & 24
\end{array} \right].
$$
We hide $t_3$ and arrive at a resultant matrix of dimension 16,
whereas the sparse resultant has degree 12.
The monic polynomial and the companion matrix are of dimension 32.
There are 16 real roots.
Four of them correspond to eigenvalues of unit geometric multiplicity,
while the rest form four groups, each corresponding to a triple eigenvalue.
For the latter the eigenvectors give us no valid information, so we
recover the values of $t_1,t_2$ by looking at the other solutions and
by relying on symmetry arguments.
The computed roots are correct to at least 7 decimal digits.
The average CPU time of the online part is $0.2$ seconds on a \sparc.

\section{Conclusion} \label{concl}

We have examined several computational aspects of
sparse elimination theory and, in particular, the
use of sparse resultant matrices for reducing root-finding to an
eigenproblem.
A general solver has been implemented based on this
approach and has been applied successfully to
fundamental problems in vision, robot kinematics and structural
biology.
These problems are of moderate size and exhibit sparse
structure as modeled by the Newton polytopes and the
mixed volume.
The efficiency and accuracy of our solver imply that
sparse elimination may be the method of choice for
such systems. 

Automating the different ways to deal with numerically unstable inputs
will improve the implementation.
For instance, clustering neighboring eigenvalues and computing the error on the
average value significantly improves accuracy.
A question of practical as well as theoretical interest is to handle the case of
repeated roots efficiently.

% \section*{Acknowledgment} 
% I would like to thank John Canny, my thesis advisor, for many insightful suggestions.

\end{document}